\documentclass[12pt,manuscript]{aastex}

\usepackage{graphicx}
\usepackage{natbib}
\usepackage[usenames,dvips]{color}
\usepackage{amssymb, amsmath, amsbsy, epsfig, epsf}
\usepackage{booktabs}

\newcommand{\mytoprule}{\specialrule{0.1em}{0.5em}{0.5em}}
\newcommand{\mybottomrule}{\specialrule{0.1em}{0.5em}{0.5em}}

\def\hb{H$\beta$}

\def\feii{\ion{Fe}{2}}
\def\oiii{[\ion{O}{3}]}

\def\rfe{$R_{4570}$}
\def\mbh{$M_{\rm BH}$}

\def\redd{$L_{\rm bol}/L_{\rm Edd}$}
\def\sgmint{$\sigma_{\rm m}$}

\def\kmps{${\rm km\,s^{-1}}$}

\slugcomment{To be submitted to {\it The Astronomical Journal}}
\shorttitle{photometric variability of NLS1-type AGNs}
\shortauthors{Ai et al.}

\begin{document}

\title{A comparative study of optical/ultraviolet variability of narrow-line Seyfert\,1 and broad-line Seyfert\,1 active galactic nuclei}

\author{
Y. L. Ai\altaffilmark{1,2},
W. Yuan\altaffilmark{3},
H. Zhou\altaffilmark{4,5,6},
T.G. Wang\altaffilmark{4,5},
X.-B. Dong\altaffilmark{4,5},
J. G. Wang\altaffilmark{2},
H. L. Lu\altaffilmark{4,5}}

\altaffiltext{1}{Department of Astronomy, Peking University, Beijing 100871, China, aiyl@pku.edu.cn}

\altaffiltext{2}{National Astronomical Observatories/Yunnan
Observatory, Chinese Academy of Sciences, Kunming, Yunnan, P.O. BOX
110, China}

\altaffiltext{3}{National Astronomical Observatories, Chinese Academy of Sciences,
Beijing, 100012, China, wmy@nao.cas.cn}

\altaffiltext{4}{Key laboratory for Research in Galaxies and Cosmology, University of Science and Technology of China, Chinese Academy of Science, Hefei, Anhui 230026, China}
\altaffiltext{5}{Department of Astronomy, University of Science and Technology of China, Hefei, Anhui 230026, China}
\altaffiltext{6}{Polar Research Institute of China,451 Jinqiao Road, Pudong, Shanghai 200136, China}

\email{aiyl@pku.edu.cn}

\begin{abstract}
The ensemble optical/ultraviolet variability of
narrow-line Seyfert\,1 (NLS1) type active galactic nuclei (AGNs) is investigated,
based on a sample selected from
the Sloan Digital Sky Survey (SDSS) Stripe-82 region
with multi-epoch photometric scanning data.
As a comparison a control sample of broad-line Seyfert\,1 (BLS1) type AGNs
is also incorporated. To quantify properly the intrinsic variation amplitudes and their uncertainties, a novel method of parametric maximum-likelihood is introduced, that has, as we argued, certain virtues over previously used methods.
The majority of NLS1-type AGNs exhibit significant variability
on timescales from about ten days to a few years with, however,
on average smaller amplitudes compared to BLS1-type AGNs.
About 20 NLS1-type AGNs showing relatively large variations are presented,
that may deserve future  monitoring observations, for instance, reverberation mapping.
The averaged structure functions of variability, constructed using the same maximum-likelihood method, show remarkable similarity in shape for the two types of AGNs on timescales longer than about 10 days, which can be approximated by a power-law or an exponential function.
This, along with other similar properties,  such as the wavelength-dependent variability, are indicative of a common dominant mechanism responsible for the long-term optical/UV variability of both NLS1- and BLS1-type AGNs. Towards the short timescales, however, there is tentative evidence that the structure function of NLS1-type AGNs continues declining, whereas that of BLS1-type AGNs flattens with some residual variability on timescales of days.  If this can be confirmed, it may suggest that an alternative mechanism, such as X-ray reprocessing, starts to become dominating in BLS1-type AGNs, but not in NLS1-, on such timescales.

\end{abstract}


\section{Introduction}

Variations of optical/ultraviolet (UV) luminosity on timescales from
weeks to years are characteristics of active galactic nuclei
(AGNs), and the study of this phenomenon is a powerful tool for
constraining models of black hole accretion.
Extensive observational investigations in the past, albeit
with mostly small samples, have revealed
dependence of the variability amplitude on various
observed quantities such as
wavelength, time-lag, luminosity, and redshift
\citep[e.g.,][]{di96,de05,meus11,wels11},
that was confirmed by using vast quasars from the Sloan Digital Sky Survey (SDSS) with repeated scans
\citep{vand04, zuo12}. With the advent of black hole mass estimation based
on single-epoch spectroscopic data on AGN, new correlations have been
suggested of the variations with black hole mass
\citep[][]{wold07,baue09}, and perhaps more fundamentally, with the
Eddington ratio as claimed independently by a few groups very recently
\citep[][]{wilh08,ai10,macl10}. The
possible dependence of variability on the fundamental physical
parameters is remarkable, as it may provide new insights into the
variability mechanisms. For instance, such a dependence favors
models of disc instability or variations in accretion rate
\citep[e.g.,][]{kawa98, li08, liu08} over other models, such as
Poisson process \citep[e.g.,][]{terl92, torr00} and gravitational
microlensing \citep[e.g.,][]{hawk93}. However, the statistical
significance of these results is not yet sufficiently high, and
further confirmation is still needed.

While considerable progresses have been
made in the past decades,
previous variability studies were focused mostly on AGNs of the
broad-line Seyfert\,1 (BLS1) type, including quasars.
Narrow-line Seyfert\,1 (NLS1) type AGNs, as a
subclass of broad emission-line AGNs (BLAGNs)
 with extreme properties,
have received little attention, however.
This may partly be ascribed to the small number of such objects
known in the past.
NLS1-type AGNs are commonly defined as having small widths of
the optical permitted emission-lines
\citep[FWHM$(H\beta)\la$ 2000 km s$^{-1}$,][]{oste85}.
In addition,  NLS1-type AGNs (see Komossa 2008 for a review)
often show strong optical Fe~II emission and weak [O~III] lines \citep[][]{gask85,good89, vero01}, a steep soft X-ray spectral slope \citep[][]{boll96, wang96},
and strong X-ray variability \citep[][]{leig99,grup04}.
Such extreme properties make them cluster at one extreme end of the strongest AGNs
correlation space, namely eigenvector 1 \citep[EV1,][]{boro92, sule00, sule07},
which is believed to be driven by the Eddington ratio (\redd).

NLS1-type AGNs are commonly thought to have
small black hole masses and thus high  accretion rates
(at a substantial fraction of or close to
the Eddington rate),
as argued extensively in the literatures \citep[e.g.,][]{mine00,sule00,coll04,grup04}.
As such, their accretion is suggested to likely proceed
via the so-called slim disk where a significant
fraction of viscosity-generated photons are trapped and advected
into the black hole, rather than the standard Shakura-Sunyaev
thin disk \citep[][]{abra88, wang03}.
As the accretion disk is responsible for the UV/optical radiation
in AGNs, different variability patterns might be anticipated for NLS1-,
compared to normal BLS1-type AGNs.

There have been a few previous variability studies
on NLS1-type AGNs, all of which were monitoring programs on relatively short
timescales (e.g. intra-night or days)  for
individual objects or small samples \citep[][]{youn99, mill00}.
The only one NLS1 with long-term (more than ten years) extensive
photometric monitoring
is Ark 564 \citep[][]{doro06}, as far as we know,
and it was found that the NLS1 has a low variability amplitude.
Based on monitoring results of a small sample,
Klimek et al. (2004)
first suggested that NLS1s, as a class, are less variable than
non-NLS1s on long timescales (weeks to years).
This is consistent with the later suggested anti-correlation with
Eddington ratio, if NLS1-type AGNs have indeed high accretion rates.

In fact, the low black hole mass scenario of NLS1-type AGNs is yet
subject to active debates (see komossa 2008 for a review),
albeit some favoring independent pieces of indirect evidence,
such as rapid X-ray variability \citep[][]{gier08,Ponti10,ai11}.
One important question concerned is whether
NLS1 AGNs conform the same relation of the size of the broad line region (BLR)
and luminosity as found in normal Seyferts and quasars (Kaspi et al. 2000).
The question arises as NLS1 AGNs  with
reverberation mapping (RM) observations
are rare, only limited to a few cases \citep[][]{pete04,kasp05}.
This situation may partly be ascribed to the lack of known NLS1-type AGNs
with large variability.
A  search of more such objects would be essential for
future RM observations of NLS1 AGNs.

We have carried out a systematic study on the optical/UV variability
of AGNs and quasars using a large sample with multi-epoch scanning data
available from the SDSS Stripe-82, including both BLS1- and NLS1-type AGNs.
As the first results, various correlations
with AGN variability have been tested and presented in \citet[][]{ai10}.
It was found that the variability amplitude is
significantly correlated with the AGN EV1 parameters, particularly with
\redd\ as the strongest, albeit large scatters \citep[][]{ai10}.
This is also manifested by the observed
averagely smaller variations of NLS1-
(having higher \redd)  compared to the BLS1-type AGNs.
However, these results are based on time-averaged variation amplitudes,
and confirmation by structure function analysis, that is a
more robust technique, is needed;
moreover, the time-lag dependence can be explored, that may shed light on
the variability difference between the two types of AGNs.

This is the second paper of our variability study
based on  the same SDSS Stripe-82 samples,
which focuses on the ensemble variability properties of NLS1-type AGNs.
The BLS1-type subsample is also analyzed in the same way
as a control sample for comparison.
These mainly concern the demography of variable objects,
wavelength dependence, and variation of the spectral shapes,
as well as the dependence on time-lag (structure function).
We also improve the measurement of variation amplitudes up on those in
the first paper (and in the literature)
by introducing the Maximum-likelihood method.
In addition, we present a full account of the data analysis, mainly
photometric calibration and reliability check, for which
only a concise description was given in the first paper given the
limited length of a letter.
The paper is organized as follows.
In Section\,2 we describe the sample selection and photometric
data analysis, and in Section\,3
the quantification of variability.
The results are presented in Section\,4,
and their implications are discussed in Section\,5,
followed by a summary in Section\,6.
We use a $\Lambda$-dominated cosmology with $H_{0} = 70~km~s^{-1}~Mpc^{-1}$,
$\Omega_{m} = 0.3$, and $\Omega_{\Lambda} = 0.7$ throughout the paper.

\section{Sample and Data}

\subsection{NLS1- and BLS1-type AGN samples}
\label{sample}
Our NLS1-type AGNs are actually taken from the sample of \citet[][]{zhou06},
which was selected from the SDSS Data Release 3 catalogs.
The procedures of the optical spectral analysis and measurement of emission line parameters were described in detail in our previous
papers \citep[][]{zhou06, dong08}.
Only objects with redshifts $z\la$ 0.8 were
included to cover the H$\beta$ and \oiii\ emission lines,
with the broad component of H$\alpha$ or H$\beta$ detected
at $>10~\sigma$ confidence level.
We further select objects for variability study using the following criteria:
\begin{itemize}
\item located within the SDSS Stripe-82 region (R.A. $>$
310$\degr$ or $<$ 59$\degr$ and -1$\degr$.25 $>$ decl. $<$
1$\degr$.25);
\item classified as
`STAR' in {\em all} of the five bands by the SDSS photometric
pipeline\footnote{The SDSS photometric pipeline differentiates
resolved (`GALAXY') and unresolved (`STAR') sources based on the
difference between PSF- and model-magnitudes, and it was found that
a simple cutoff at 0.145 mag works at the 95\% confidence level for
sources with $r\leq21$ mag \citep[][]{stou02}.}  to
minimize possible contamination from host galaxy starlight;
\item radio-quiet or not detected in the FIRST radio survey\footnote{Defined as $R_{1.4}<1$, where the radio-loudness
$R_{1.4}$ is defined as the logarithm of the flux density ratio of
radio emission at 1.4 GHz to the optical emission at the SDSS
$g$-band.}, as constrained from the FIRST \citep[Faint Images of the Radio
Sky at 20 cm survey,][]{beck95} radio survey. This is to
eliminate possible `contamination' from blazar-type jet emission,
which originates from a distinct (non-thermal) radiation process and is often
variable \citep[][]{zhou03, zhou07, yuan08, abdo09, liu10, jiang12}.
\end{itemize}

\begin{figure}
\centering\includegraphics[scale=0.8]{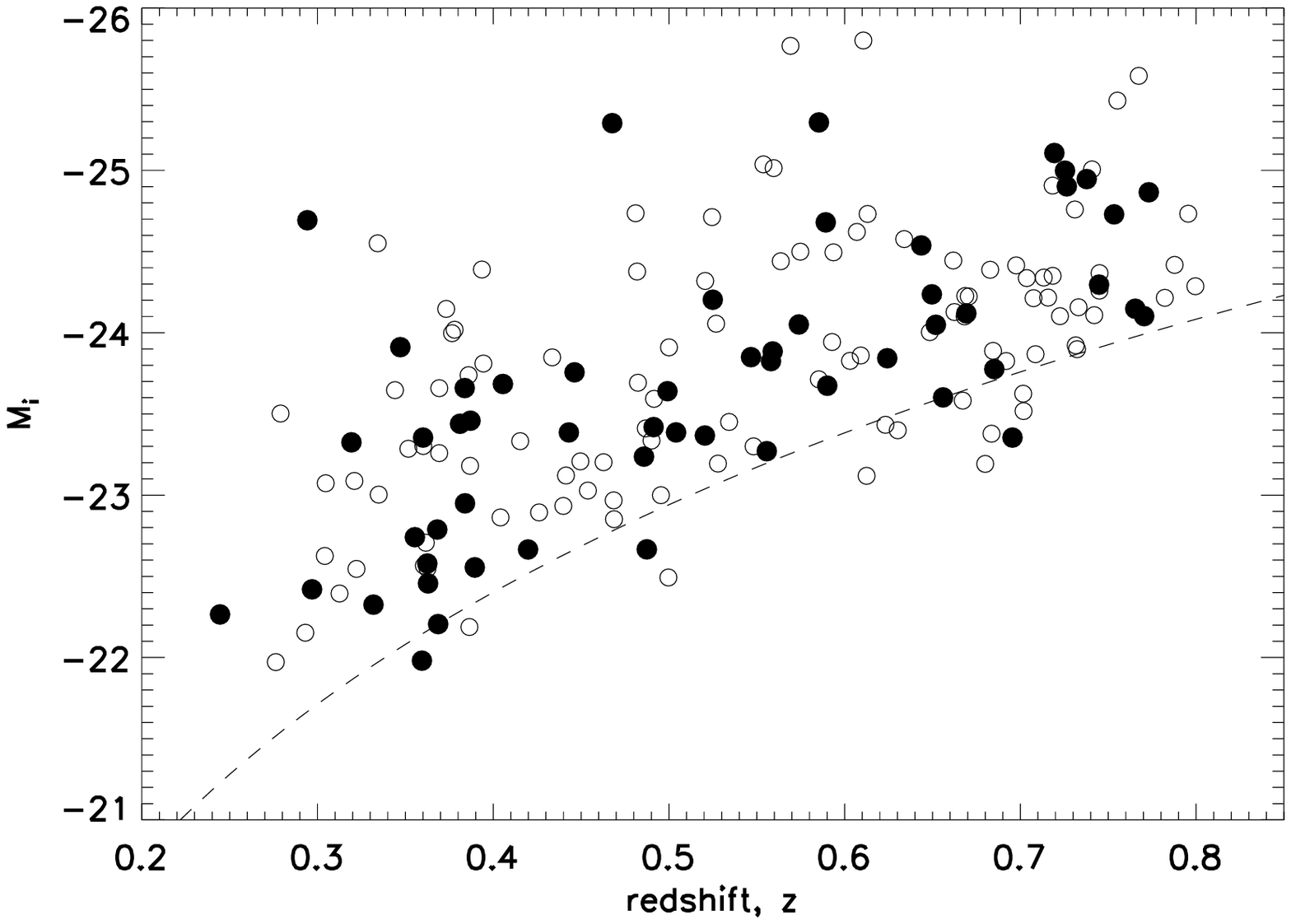}
\caption{Redshift and $i$-band luminosity distribution of our
NLS1-type AGN (filled circles) and BLS1-type (open circles) samples.
The dashed curve represents the limiting magnitude for optically-selected
quasars in the SDSS ($m_{i}=19.1$).
\label{Lum_Z}}
\end{figure}

The above selections result in 58 NLS1-type AGNs. For a comparative study
we also select a BLS1-type AGN sample in the Stripe-82 region, i.e., with the broad H$\alpha$ or
H$\beta$ linewidth (full width at half maximum, FWHM) greater than 2,200\,\kmps, the dividing line used
to separate NLS1- and BLS1-type AGNs following our previous
work (Zhou et al. 2006, see also Gelbrod et al. 2009).
The sample selections
and spectral analysis
are performed in the same way as for
NLS1-type AGNs above, and is described in \citet[][]{dong08}.
The same criteria as above are also applied to select star-like, radio-quiet objects.
Furthermore, To ensure that the two samples match each other on the redshift-luminosity distribution, a subsample of the resulting BLS1-type AGNs is randomly extracted in a way that they mimic
the $z-M_{\rm i}$ distribution of the NLS1-type sample,
where $M_{\rm i}$ is i-band absolute magnitude\footnote{A power law continuum $f_{\lambda}
\propto \lambda^{\alpha_{\lambda}}$ with a slope of
$\alpha_{\lambda}= -1.5$ is assumed in calculating the i-band
absolute magnitude. The observed magnitudes have been corrected for
the Galactic extinction \citep[][]{schl98}.}.
In this process we try to retain the size of the NLS1-type sample
(with only a few outliers discarded) and prune the BLS1-type sample
with a BLS1-type to NLS1-type ratio consistent with 2:1.
This results in two final working samples of
55 NLS1-type and 108 BLS1-type AGNs, which are
statistically compatible with each other on the
redshift--luminosity plane (Figure\,\ref{Lum_Z}).
The 2-D Kolmogorov-Smirnov test \citep[][]{press92}
yields a chance probability of 0.43 that the two samples have the same
$z-M_{\rm i}$ distribution.

\begin{figure}
\centering\includegraphics[scale=0.8]{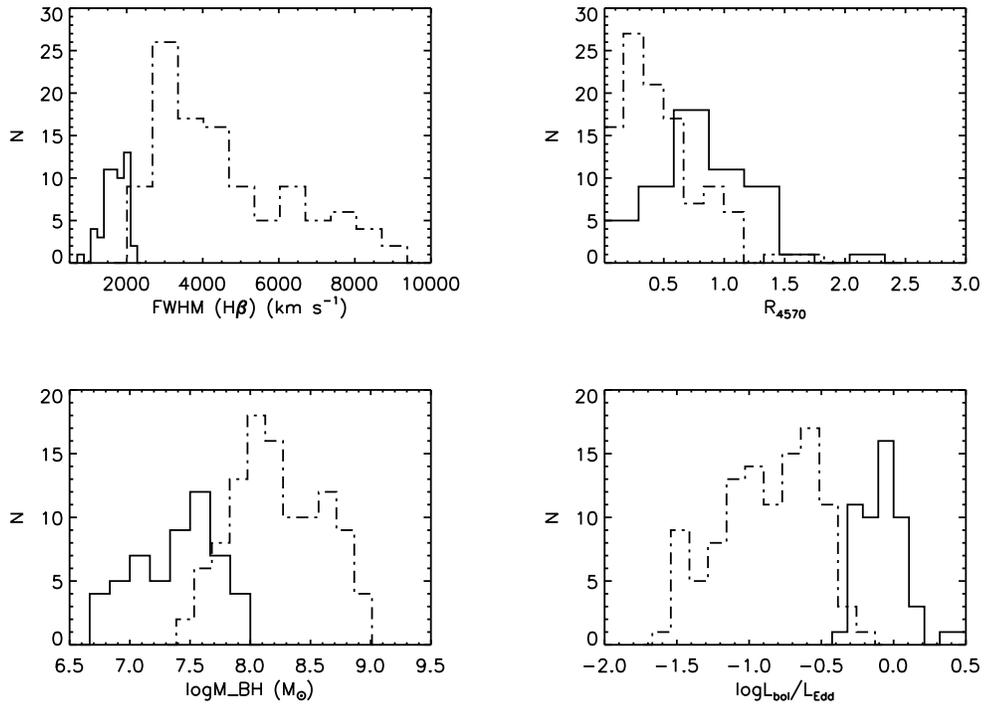}
\caption{Distributions of the \hb\ linewidth, relative strength
of \feii\ emission, black hole mass and
Eddington ratio for the NSL1- (solid) and BLS1-type (dash-dotted) AGNs.
\label{sample_hist}}
\end{figure}

The two working samples can be considered as optically and homogeneously selected,
with reliably measured continuum and emission line parameters \citep[see][]{zhou06, dong08}.
They consist of mostly quasars with $M_{i}<-23$\,mag and $z\simeq$ 0.3--0.8.
All of the NLS1-type objects meet the conventional extra \oiii/\hb$<3$ criterion for
NLS1s (see zhou et al. 2006, for details).
They also show, compared to the BLS1 sample, other properties characteristic of NLS1s, e.g.,
stronger Fe~II emission, smaller
black hole masses and higher Eddington ratios (Figure\,\ref{sample_hist}).
The strength of the Fe~II emission lines is measured by the intensity ratios to \hb,
i.e., \rfe $\equiv$\feii ($\lambda \lambda 4434-4684) / \rm H\beta^b$
(where \feii\ ($\lambda\lambda4434-4684)$ denotes the \feii\
multiplets flux integrated over 4434--4684\,\AA, and H$\beta^b$ the flux of the broad component of \hb, see Zhou et al. 2006 for details).
The black hole masses, \mbh, are estimated using the formalism given by \citet[][]{gree05} from
the broad \hb\ FWHM and the 5100\,\AA\ luminosity ($L_{5100}$), which are taken
from \citet[][]{zhou06}, or measured in the same way for the BLS1 sample.
To calculate the Eddington ratio, \redd, the
bolometric luminosities are estimated as $L_{\rm bol}=9L_{5100}$ \citep[][]{elvi94}.

\subsection{Data and photometric calibration}

The SDSS imaged the sky
in five broad photometric bands simultaneously,
namely, $u, g, r, i$ and $z$ \citep[][]{gunn06, fuku96}.
The integration time is 54.1s in each band and the limiting magnitude reaches $\sim 23$ mag in the $r$ band.
The photometric system is based on the AB-system with
zero-point uncertainty of $\sim$0.02--0.03 mag \citep[][]{smit02, fuku96, Ivez04}.
The astrometric accuracy is better than $0.\arcsec1$ for
sources brighter than 20.5 mag in $r$ band \citep[][]{pier03}.
We use the point-spread function (PSF) magnitudes in this work.

The SDSS Stripe-82 survey covers the region from $\alpha = 59\degr$ to $310\degr$ and
$\delta = -1.25\degr$ to 1.25$\degr$. During the SDSS-I phase ($\sim$ 2000--2005) the
region was repeatedly observed and the central
part of the stripe has been scanned at a cadence of typically 10--20 times \citep[][]{adel07, sesa07}.
These observations were performed under generally photometric conditions and the data were well
calibrated using the PHOTO software pipeline \citep[][]{lupt02}.
This region was later repeatedly scanned over the course of three 3-month campaigns (Sep.--Nov.)
in the successive three years in 2005--2007, known as the SDSS Supernova (SN) Survey.
In this work we use the photometric data acquired during both the SDSS--I phase from Data Release 5 \citep[DR5,][]{adel07}
and the SN survey during 2005 (SN-2005).

Observations in the SN survey were sometimes performed in non-photometric conditions.
At the time when this work was started only the un-calibrated source catalogs were available.
Thus possible photometric zero-point offsets in the SN survey data need to be determined, by calibrating against the DR5 magnitudes.
To do this, we use stars in the same fields as `standards', assuming that the vast majority of stars do not vary.
The detailed procedures of re-calibration are described in Appendix \ref{sect:calibration}.
The resulting overall (systematic and statistical) photometric errors
of the calibrated SN survey magnitudes have a median of $\approx0.03$\,mag for the $g$, $r$, and $i$ bands, and
$\approx 0.04$\,mag for the $u$ and $z$ bands, which are comparable to those of the SDSS-I DR5 data.
To check the reliability of our photometric calibration,
we examine the calibrated SN survey data
of 14 Landolt photometric standard stars \citep[][]{land92} locating in Stripe-82;
none  of them is found to show detectable variability.

\section{Quantification of variations and their uncertainties}
\label{measurement}

We construct light curves for each object
in the five SDSS bands separately using the
DR5 and the re-calibrated SN-2005 Survey data.
Objects observed in the same frames observed at different epochs
are matched by using a
matching radius of 1\arcsec.
To eliminate effects of observation condition,
only frames with good image quality
(flagged as ¡°good¡± or ¡°acceptable¡±) are used.
For each object there are typically $\sim$27 observations
(about 14 from the SN-2005 survey)
spanning $\sim$5 years.
It should be noted that all sample objects have similar
sampling patterns, mostly being observed repeatedly in three months (Sep.--Nov.) each year during 2000--2005.

First we use the chi-square method to test the significance of variability for
individual objects against the null hypothesis of no
variation. The statistic  $\chi^2$ \citep[e.g.,][]{bevi92} is defined as
\begin{equation}\label{eq1}
\chi^2 = \sum\limits_{i=1}^{N}{\frac{(m_{i}-\langle m
\rangle)^2}{\xi_{i}^{2}}},
\end{equation}
where $m_{i}$ is the  magnitude of the {\it i}-th measurement with
error $\xi_{i}$ (the overall error including both statistical and systematic errors),
and  $\langle m \rangle$ the weighted mean of a total of N measurements.
We consider a source to be variable only if the probability level against
the null hypothesis
$P<0.1\%$.

The amplitude of variability of an object is commonly measured by the
width of its magnitude distribution.
However, the magnitude distribution {\em as observed} does not represent the {\em intrinsic} variability,
but rather is broadened by the effect of measurement errors.
To assess the intrinsic magnitude distribution, we first take the common practice as used in the
literature \citep[e.g.,][]{vand04, sesa07}, by estimating the contribution of measurement errors to the observed scatter (root-mean-square, rms)
and then subtract it from the latter (See Eq. B2 in Appendix\,\ref{sect:method} for a detailed account).
This estimation is most accurate when the magnitude uncertainties
$\xi_{i}$ have the same or very similar  values.
However, this may not be the case for the SDSS SN survey data that taken
under variance of observing conditions.
Moreover, for such estimation and its variants,
it is difficult to quantify the uncertainties of the
estimated intrinsic variability amplitudes.

\begin{figure}
\includegraphics[scale=0.8]{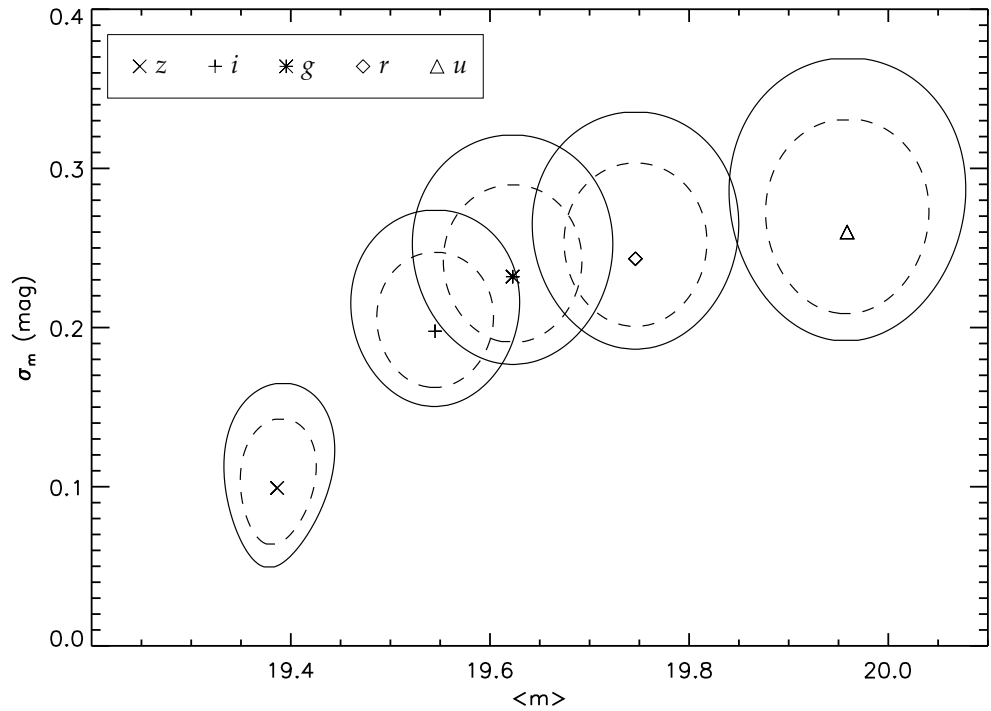}
\centering \caption{Example of the best estimates and confidence contours of the
mean $\langle m \rangle$ and dispersion $\sigma_{m}$ of the
intrinsic magnitude distribution at the $90\%$ and $60\%$
confidence levels in $ugriz$ bands for one NLS1-type AGN. \label{maxi_example}}
\end{figure}

Alternatively, we introduce a parametric maximum-likelihood method,
in which the intrinsic magnitude distribution of an object is parameterized as a Gaussian with a
mean $\langle m \rangle$ and a standard deviation $\sigma_{m}$; the standard deviation $\sigma_{m}$ is used as a measure
of the amplitude of intrinsic variability. This method can take into account the uncertainty of each
individual measurement (assumed to be Gaussian distributed). Moreover,
it can also be used to quantify the confidence intervals for
interesting parameters, such as $\sigma_{m}$, (Figure\,\ref{maxi_example}).
The method had been used in a similar way in the literature
to derive the intrinsic distribution
of observables which suffer from measurement uncertainties \citep[e.g.,][]{macc88}.
The description of this method and a comparison with the one above are
in Appendix\,\ref{sect:method}. The two methods are found to be in excellent agreement.
Hence in this paper we use the amplitudes derived from the
maximum-likelihood method since their confidence intervals can be estimated for each object.

\section{Results}
\subsection{Demography of Variability}
\label{demography_variability}

\begin{figure}
\includegraphics[scale=0.8]{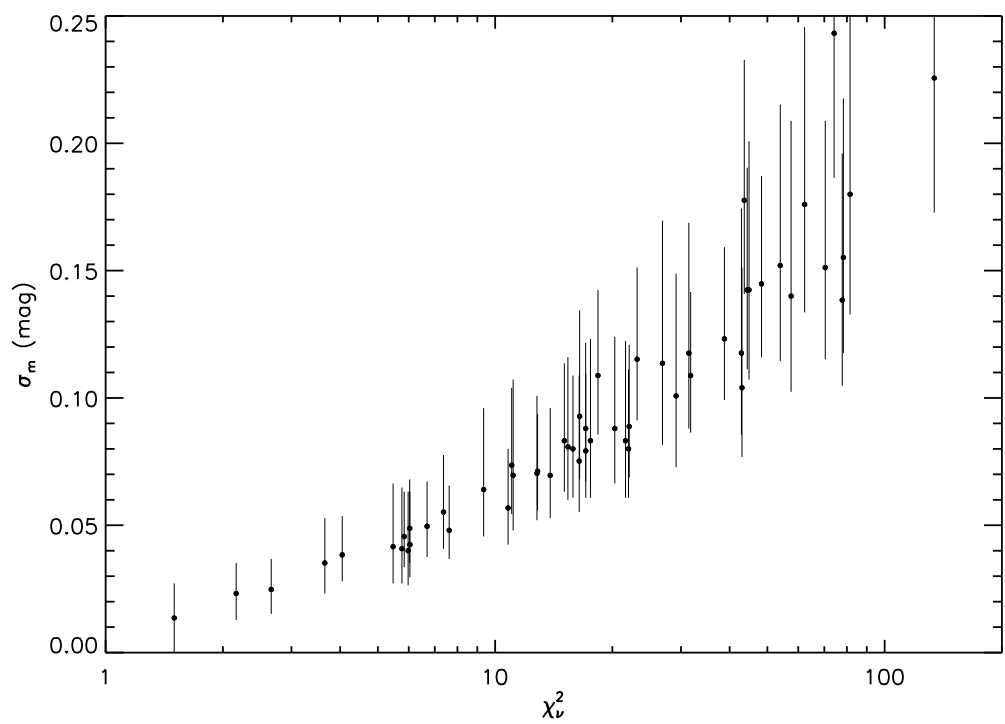}
\centering \caption{The $r$-band variability amplitudes versus reduced
chi-squares for NLS1-type AGNs. \label{chis_amplit}}
\end{figure}

Here we quantify the demography of the variability.
We consider an object to be variable if the $\chi^2$
null-variability probability $P<0.1\%$, which corresponds to
$\chi^2_{\nu} \geq 3$ for a degree of freedom of 27,
roughly the mean number of data sampling of our objects.
Table\,\ref{var_fraction} lists the fraction of variable objects
for both the NLS1- and BLS1-type AGN samples.
We find that the vast majority of NLS1-type AGNs show variations on
timescales of years at the significance level $P<0.1\%$.
Most of the variable objects varied at a level,
\sgmint\ $>$ 0.05\,mag, larger than the typical photometric errors.
This is clearly demonstrated
in Figure\,\ref{chis_amplit}, which shows the reduced chi-square
versus the variability amplitude
in the $r$ band as example. We can see that for
NLS1-type AGNs the intrinsic variability amplitude \sgmint\ ranges from $\sim0.25$ mag down to a few per cent
at $\chi^2_{\nu} = 3$; the latter can be considered as the sensitivity of the
variability detection limited by the current data quality.
As can be seen, the optical/UV emission of NLS1-type AGNs are indeed variable on timescales of years,
with mean variability amplitudes around 0.07--0.11 mag, depending on bands.

\begin{figure}
\centering \includegraphics[scale=0.8]{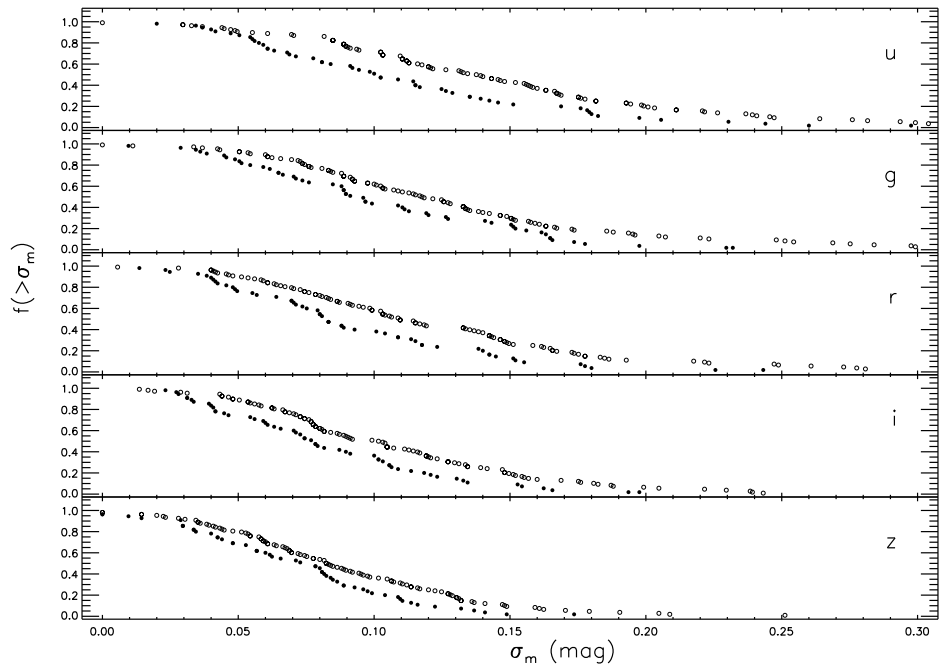}
\caption{Accumulative distribution of the variability amplitudes, \sgmint, for NLS1- (filled circles) and BLS1-type AGNs (open circles) in the five SDSS photometric bands.
The x-axis ranges truncate at 0.3 mag for clarity. \label{var_hist_rest}}
\end{figure}

The accumulative distributions of the variability amplitudes \sgmint\ in the five SDSS bands
are shown in Figure\,\ref{var_hist_rest}. Several results can be inferred
immediately. First, most of both the NLS1- and BLS1-type AGNs are variable
on timescales of years, and there is a large scatter,  stretching
over about one order of magnitude, in the variability magnitude.
Second, the NLS1- tend to vary with smaller amplitude compared to the BLS1-type AGNs, and this trend
is most significant in the blue bands.
This result is in line with our previously found correlations involving variability,
i.e., the narrower line width, the stronger \feii\ emission, the higher Eddington ratio, and the
smaller variability \citep[][]{ai10}.
Third, For both NLS1- and BLS1-type objects the variations are larger in the bluer band
than the redder band; this is discussed in detail in Section\,4.2.

\begin{figure}
\centering \includegraphics[scale=1]{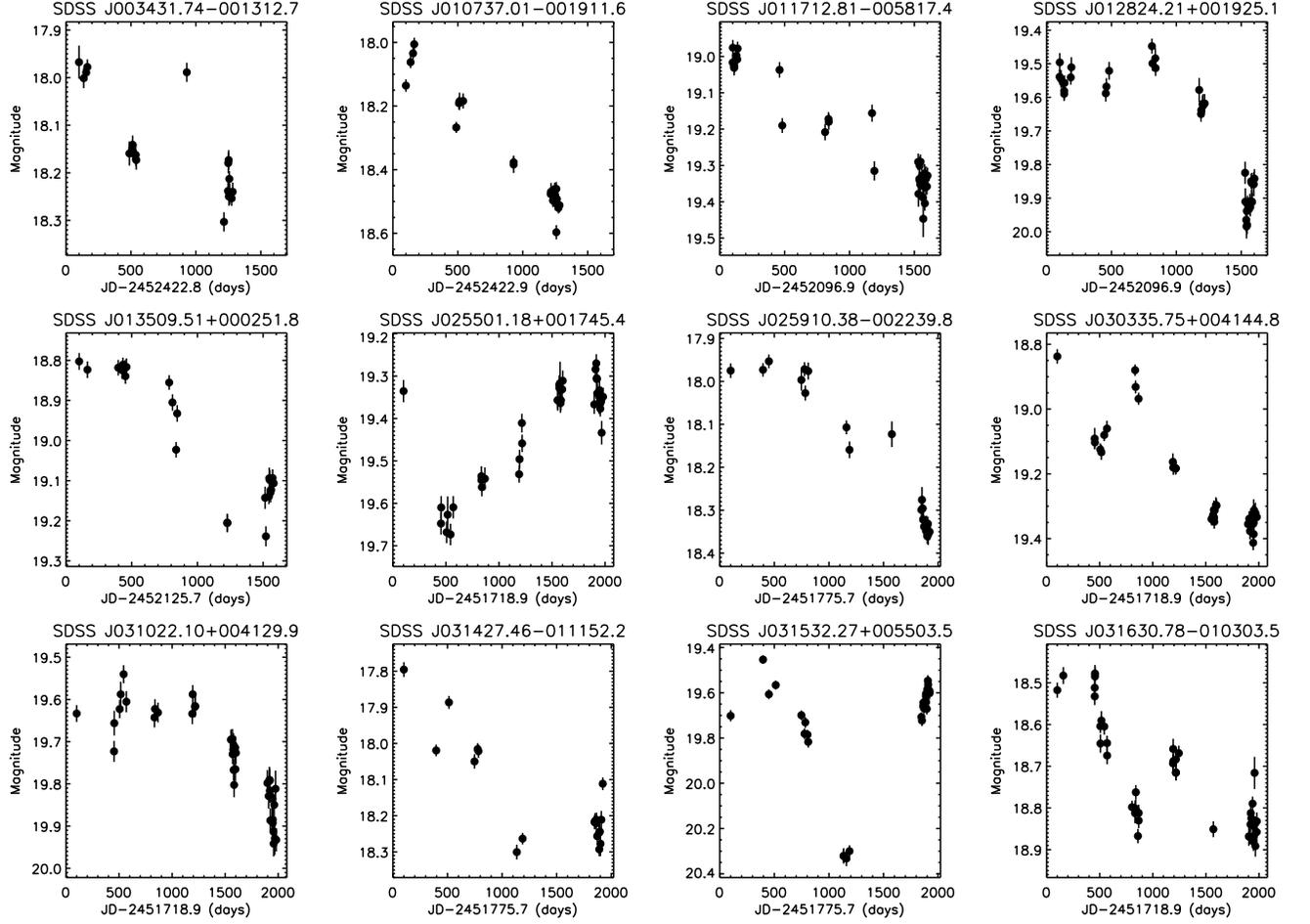}
\caption{Light curves of the NLS1-type AGNs in $r$ band with \sgmint$>0.1$\,mag. \label{lightc_22}}
\end{figure}

Variable NLS1-type AGNs with relatively large amplitude are of
particular interest and suitable for future spectroscopic monitoring and reverberation mapping observations.
We list objects with $r$ band \sgmint$>0.1$\,mag in Table\,\ref{NLS1_examp}, including some of the
continuum and emission line parameters, as taken from \citet[][]{zhou06}.
We note that most of these are typical NLS1-type AGNs with strong \feii\ emission.
Their light curves are shown in Figure\,\ref{lightc_22}.

\subsection{Wavelength dependence and spectra variability}
\label{wavelength_depend}

\begin{figure}
\centering \includegraphics[scale=0.8]{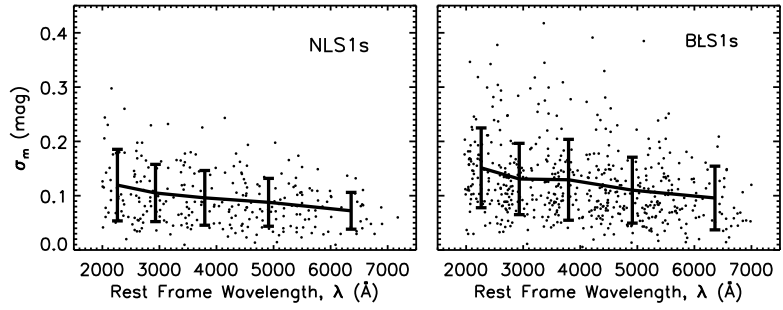}
\vspace{-2cm}
\caption{Variability amplitude as a function of rest-frame
wavelength for NLS1- (left) and BLS1-type AGNs (right). The mean value in each bin is
also shown with the error bar representing the standard deviation of the distribution in each bin.
\label{var_rest_wave}}
\end{figure}

Wavelength-dependent variability has long been found for BLAGN and QSO samples \citep[e.g.,][]{vand04}.
Here we examine this property for NLS1-type AGNs using our large data set.
Due to redshift, a given filter samples different wavelength ranges for objects with different redshift.
We approximate the derived amplitude \sgmint\ for a filter band as that at its effective
wavelength $\lambda_{0}$. For an object of redshift $z$, $\lambda_{0}$ corresponds to a wavelength
$\lambda=\lambda_{0}/(1+z)$ in the object's rest frame. Thus for each object the variability \sgmint\
at five rest-frame wavelengths are sampled corresponding to the five SDSS bands.
we divide the overall rest wavelength range sampled here, 1900$\sim$7100\,\AA, into five bins of
equal bin-size in $\log\lambda$£¬\footnote{The rest wavelength bin, in \AA, are 1900, 2500, 3300, 4200, 5500, 7100.}.
The relationship of amplitude \sgmint\ versus rest-frame effective wavelength $\lambda$
is shown in Figure\,\ref{var_rest_wave}. The mean value and scatter (standard deviation) of the sigma distribution in each wavelength bin is also overplotted. The result shows clearly a trend of increasing variability amplitude
moving towards short wavelength for NLS1-type AGNs,
similar to BLS1-type. It shows clearly that at a given wavelength NLS1-type AGNs have systematically lower variation amplitudes
compared to BLS1-type, albeit large scatters. We test the
significance of the differences in the \sgmint\ distributions
between the two samples using the K-S test, yielding the chance
probability for the same distribution as 1.2\%, 9.7\%, 2.9\% for the
shortest three wavelength bins, respectively.

\begin{figure}
\centering \includegraphics[scale=0.8]{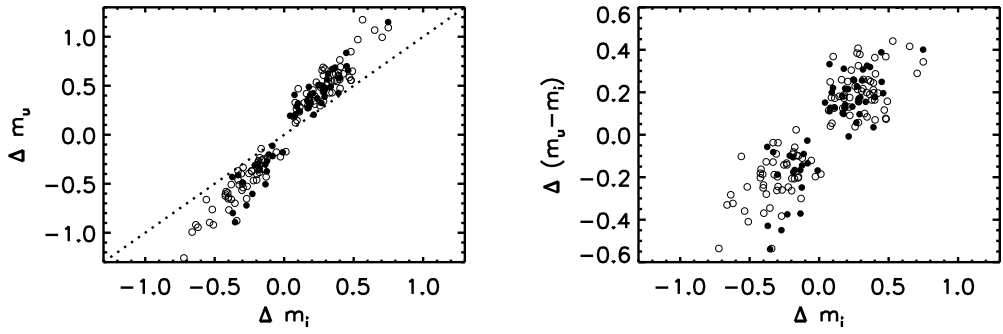}
\vspace{-2cm}
\caption{left: Comparison of the maximum magnitude differences
between the $u$ and $i$ bands
for each NLS1- (filled circles) and BLS1-type AGNs (open circles); right:
the changes in the color versus the maximum magnitude differences.
\label{var_color}}
\end{figure}

The trend in Figure\,\ref{var_rest_wave} is in statistical sense only, and it
would be interesting to examine whether it also holds
for variability in individual objects.
For each object, we estimate the maximum magnitude difference over the entire light curve
in the blue ($u$) and red ($i$) bands, respectively,
defined as $\Delta m=m_{\rm t2}-m_{\rm t1}$, where t1 and t2 are the epochs of the
two observations and t2$>$t1.
The result is shown in the left panel of Figure\,\ref{var_color}; the
dotted line indicats the 1:1 relationship where the
variations in the two bands are identical in both the sign
and amplitude. Two conclusions can be inferred from the figure.
First, $\Delta m$ in both the blue and red bands always keep the
same signs, meaning that the two bands vary (fading or brightening)
coordinately; That is to say, the optical/UV continuum level moving up and
down systematically, rather than seesawing around. Second, for all
but a few objects the variations in the blue are indeed larger than
those in the red band. This suggests that, even in individual AGNs, the brightness variation tends to go hand-in-hand
with variation of the continuum slope, with larger variability in
blue than in red. This is the same for both NLS1- and BLS1-type AGNs.

A direct consequence of such wavelength-dependent variability is
that the optical/UV continuum becomes redder (bluer) when the
overall continuum level becomes fainter (brighter). This can be
demonstrated explicitly in the right panel of Figure\,\ref{var_color}, where the
changes in the color ($\Delta(m_{u}-m_{i})$) are plotted against the changes of magnitude.
As can be seen, for all but a few objects,
of both BLS1-
and NLS1-type AGNs, the spectra redden ($\Delta(m_{u}-m_{i})>0$) while objects fade
away ($\Delta m>0$), and vice verse. Furthermore, there is a strong
correlation between the change of color ($\Delta(m_{u}-m_{i})$) and the
change of magnitude ($r_{\rm s}$=0.79, $P \leq 10^{-5}$, the Spearman rank correlation test).
This implies that the larger variation of the continuum level is, the more change of the
spectral shape in general.
We note that, regarding the dependence of variability on wavelength and spectral shape, NLS1- and BLS1-type
AGNs show very similar behavior.

\subsection{Time dependence of variability}

To characterize the dependence of variation on time lags, structure function is widely used \citep[see, e.g.,][]{coll01, vand04}. In this paper we construct the structure function with maximum-likelihood method, as described in Section\,\ref{measurement}, to parameterize the intrinsic variability amplitude for the ensembles instead of individuals. Firstly, for each object, we calculate the magnitude difference between
any two observations separated by an interval, $\Delta\tau$, in the object rest frame, $\Delta m=m(t)-m(t-\Delta\tau)$.
The error of $\Delta m$ is estimated using error propagation.
The combined data points for all objects in the samples are then grouped into various bins according to time lags.
The time lag bins are divided in a way that there are at least 600 data points in each bin.
In each bin of time lag, the mean of the $\Delta m$ distribution is expected to be around zero for a large number of data points, since objects brighten or dim randomly and evenly in the statistical sense. The dispersion of $\Delta m$ distribution, in fact, reflects the degree of variations (at the given time lag). We take the intrinsic dispersion of the $\Delta m$ distribution in each bin as a measure of the `averaged' amplitude of variability, i.e., the structure function. The intrinsic dispersion can be obtain with the assumption that $\Delta m$ distribution in each bin can be approximated by a Gaussian, which is verified in the majority of the time lag bins. The uncertainties of the intrinsic dispersion can also be obtained at the same time. In this way we construct the structure function with maximum-likelihood method.

\begin{figure}
\centering \includegraphics[scale=0.8]{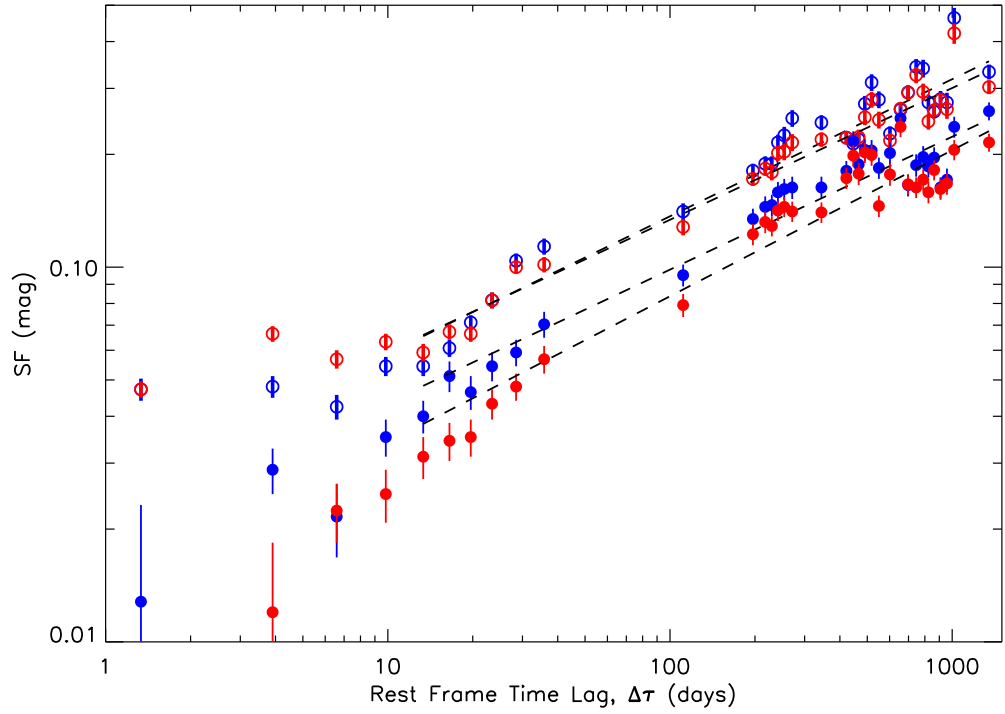}
\caption{Structure functions of the NLS1- (filled circles) and BLS1-type AGNs (open circles) in $g$ (blue) and $r$ (red) bands. The errors are at $90\%$ confidence level. The dashed lines are the power law model fittings to data at $\Delta\tau>$10 days.
\label{SF_maxilm}}
\end{figure}

We first construct the structure function for NLS1-type and BLS1-type AGNs in each SDSS photometric band, for the sake of easy comparisons with previous results of BLAGNs and quasars. The derived structure functions are shown in Figure\,\ref{SF_maxilm}, in only $g$ and $r$ bands for clarity. We find that the variability amplitude of NLS1-type AGNs, similar to BLS1-type, does increase with time lag from days to years in all of the SDSS bands.
At a given timescale, NLS1- have systematically lower variation amplitudes than BLS1-type AGNs.
A simple power law model is used to characterize the structure function at time lags greater than 10 days,
\begin{equation}
SF(\Delta\tau) = \left(\frac{\Delta\tau}{\Delta\tau_{0}}\right)^{\beta}
\end{equation}
where $\Delta\tau_{0}$ and $\beta$ are to be determined.
The fitting results are given in Table\,\ref{SF_model_fits} and Figure\,\ref{SF_maxilm}.
It can be seen that a power law model can represent the structure function reasonably well.
The inferred power law slopes for NLS1-type AGNs range
from 0.32$\pm$0.02 to 0.40$\pm$0.01, mostly consistent with those of the BLS1s within the uncertainties.

\begin{figure}
\centering \includegraphics[scale=0.8]{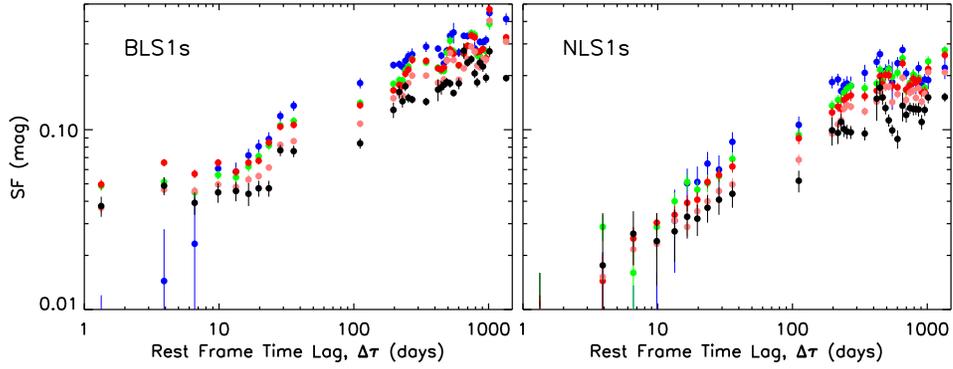}
\caption{Structure functions of the BLS1- (left ) and NLS1-type AGNs (right) in rest-frame wavelength bins of
$\Delta\lambda (1900-2500\,\AA)$ (blue), $\Delta\lambda (2500-3300\,\AA)$ (green), $\Delta\lambda (3300-4200\,\AA)$ (red), $\Delta\lambda (4200-5500\,\AA)$ (pink), and $\Delta\lambda (5500-7100\,\AA)$ (black).
\label{SF_maxilm_rest}}
\end{figure}

Considering the redshift effect, a more physically meaningful approach is to construct the
structure function in rest-frame wavelength bins instead of in photometric bands.
We divide the whole rest wavelength range into five bins with equal bin-size in logarithm, as described in Section\,\ref{wavelength_depend}. The constructed structure functions for both NLS1-type AGNs and BLS1- in each bin are shown in Figure\,\ref{SF_maxilm_rest}. As expected, the time lag dependence of variability presents in all the bins.
Wavelength dependence variability is also clearly present in almost all of the time lag bins, which is consistent with what was found above and also in previous work for BLAGNs. Again, the two types of AGNs show remarkable similarity in the shape of their structure function on timescales longer than about 10 days.

On shorter timescales (below about 10 days), the structure functions of the two types seem to differ, however. The structure function of NLS1-type AGNs continues to drop, whereas that of BLS1 seems to flatten towards short time lags, with some residual variability of several percent. This is the case for all the wavelength bins except for the shortest wavelength, for which the photometric uncertainties are the largest.  Also we note that the wavelength-dependence effect becomes weak or even vanishes on such timescales. This may indicate a different mechanism of variability starts to become dominating. However, this is not the case for NLS1-type AGNs, of which the structure function continues to drop towards short time lags. Given the sparse sampling of the data used here, further confirmation is needed to verify this trend.

\begin{figure}
\centering \includegraphics[scale=0.8]{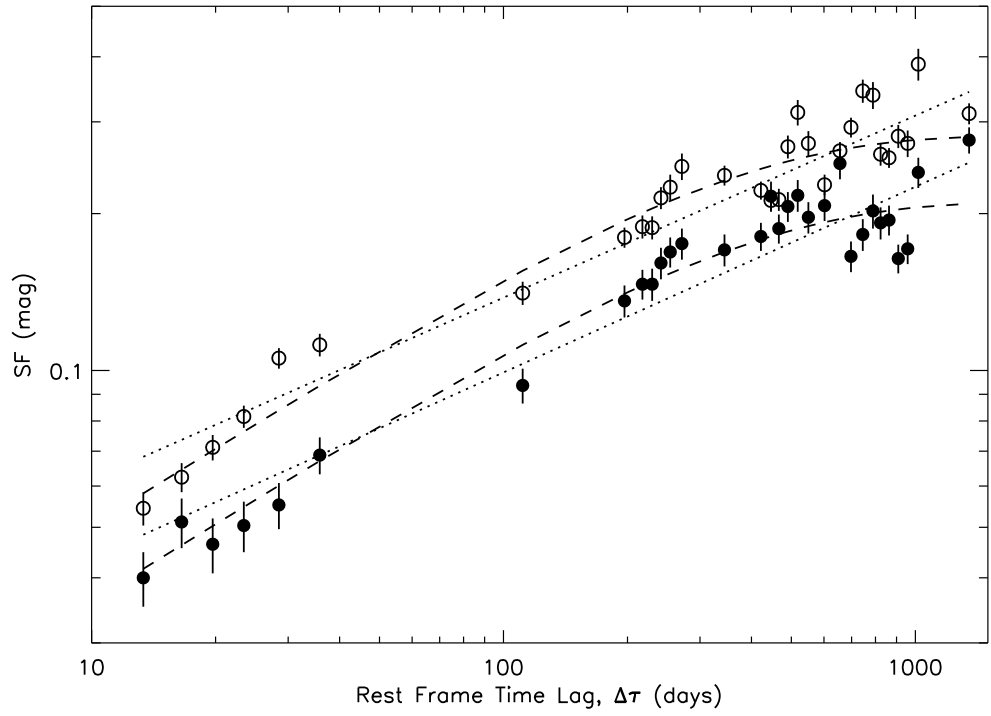}
\caption{Structure functions of the NLS1-type AGNs (filled circles) and BLS1-type AGNs (open circles) in rest wavelength bin, 2500-3300\,\AA. The fittings with power law (dotted) and exponential model(dashed) are also presented.
\label{SF_model}}
\end{figure}

The structure functions in various rest frame wavelength bins are also fitted with a power law model within the time lag range of greater than 10 days.
The fitted curves are shown in Figure\,\ref{SF_model} for the bin of 2500-3300\,$\AA$ as an example, and the results are given in Table\,\ref{SF_model_fits}.  Alternatively, an exponential model \citep[][]{macl10} is also used
\begin{equation}
SF(\tau) =SF_{\infty}(1-e^{-\mid\Delta t\mid/\tau})^{1/2}
\end{equation}
where SF$_{\infty}$ and $\tau$, the time scales, are to be determined. With inferred SF$_{\infty}$ we can estimate the long-term standard deviation of the variability as $0.5(SF_{\infty})^2$.
It can be seen that, although both the functions can approximate the observed structure functions reasonably well (at $\tau >$10 days), we note that the exponential function gives a somewhat better description than the power law for the NLS1 sample, based on the $\chi^2_{\nu}$ statistic. This may be ascribed to a possible flattening of the structure function at the largest time lags, which appears to be more significant for the NLS1- than for BLS1-type AGNs (see Figure\,\ref{SF_maxilm_rest}, Figures\,\ref{SF_model}). However, the time lags in the current data set are not long enough to draw the conclusion.
The inferred variability timescales are around one year and increase towards shorter rest-frame wavelength bins.
The values of SF$_{\infty}$, i.e., the maximum variation amplitude at the longest possible time lags, as expected, are larger at shorter wavelength than longer, and larger for BLS1- than NLS1-type AGNs in all the rest wavelength bins.

\section{Discussion}

\subsection{Implications to the optical/UV variability of NLS1s}

The similarities found between NLS1- and BLS1-type AGNs in most of the properties
of optical/UV variability, except for systematically smaller amplitudes of NLS1s, indicate that their long-term variability must
be driven by the same mechanism.
Recently there are compelling evidences that the variability is intrinsic to AGN activity, specifically the detected correlations of optical/UV variabilities
with BH masses and/or Eddington ratios, as investigated in our previous study \citep[][]{ai10} and other research \citep[][]{wold07, wilh08, baue09, zuo12}.
The variability we observed may be caused by
variations in the accretion rates
[see Gaskell (2008) for a different view, however],
or some kind of (local)
disturbance and its propagation in the disk \citep[][]{czer06}.
Assuming the first case,
\citet[][]{li08} proposed a simple model of global change of the disk structure
due to variation in accretion rates based
on the standard disk, which explained the previously observed
\mbh\ and luminosity dependence of variability.
In fact, this model can also reproduce qualitatively
an inverse amplitude--\redd\ relation (Li, private communication),
which is \mbh--dependent, though a quantitative comparison with our data
is hampered due to the small sample size.

Let's consider the case of local instability scenario.
From the simple standard disk model \citep[][]{shak73},
the emission at a given radius R is dominated by  a local blackbody
with $T_{\rm eff}$ peaking at a wavelength $\lambda$,
where
\[h c /\lambda\simeq 2.8kT_{\rm eff} .
\]
The radiation is balanced by
the energy generated per unit disk area at the radius $R$,
\[
\sigma T_{\rm eff}^4\simeq 3GM_{\rm bh} \dot{M} /(8\pi R^3),
\]
where $\dot{M}$ is the mass accretion rate.
Combining the two relations we find the radius $R$
where the local blackbody emission is peaked at $\lambda$ is,
in units of the Schwarzschild radius $R_{\rm Sch}$,
$r\equiv R/R_{\rm Sch} \sim (\dot{m}/M_{\rm bh})^{1/3} \lambda^{4/3}$,
where $\dot{m}$ the mass accretion rate in units of the Eddington rate.
This means that the scaled radius $r$,
which dominates the emission in a given bandpass, enlarges with increasing
Eddington ratio.
If the disturbance is generated in the inner part
of the disk and propagating outward,
it would attenuate in the course of propagation.
This would account for the ensemble weaker variability of NLS1-type AGNs compared to BLS1-type AGNs, since the radius $r$ responsible for the emission in the same bandpass
is shifted further out as the Eddington ratio increases.
On the contrary, if the disturbance is generated in the outer disk
and is propagating inward, the disturbance must be amplified instead.

Reprocessing of the X-ray emissions, originating close to the central black hole and known with strong and rapid variability, into the optical/UV bands also can cause the optical/UV variability in AGNs \citep[e.g.,][]{czer06, gask08}.
Yet the multi-bands variability correlation studies in individual AGNs indicate that the dominant mechanism for the long-term AGN optical/UV variability can not be reprocessing of variable X-ray emission \citep[e.g.,][]{arev08, arev09}.
Our result about the statistically weaker variability of NLS1- than BLS1-type AGNs in optical/UV bands supports the
argument, since NLS1-type AGNs often show strong X-ray variability on both short and long timescales \citep[e.g.,][]{papa04}.
However, on timescales of less than 10 days, it is likely that in this regime the X-ray reprocessing starts to take over and dominate the optical/UV variability, as the relevant timescales involved in this process are generally very short \citep[][]{czer06}.

We find that the variability component at short-term timescales seems to behave differently from the long-term one in that the wavelength-dependence is absent, or at least weak. And the structure function of BLS1-type AGNs begins to flatten at time lags less than 10 days. Interestingly, this flattening is not seen in the structure functions of NLS1s, which remain declining in the power law form. The effect of contaminations from emission line variability is excluded since the emission lines are mostly not variable on timescales of days \citep[][]{vest05, wilh05}.
We propose that such a difference may be explained in terms of the X-ray reprocessing scenario. As suggested in \citet[][]{uttl06}, the interaction between optical/UV and X-ray variability maybe different between NLS1- and BLS1-type AGNs. The optical/UV emission regions for NLS1-type AGNs are outer in the accretion disk compared to BLS1-, which subtend only a smaller solid angle as seen from the X-ray sources; while in BLS1-type AGNs the optical/UV emissions originate much closer to the X-ray emitting region, which means much tighter interactions between X-ray and optical/UV emissions. Moreover, the X-ray emission in NLS1-type AGNs  appears to be generally weaker and softer than BLS1-type AGNs \citep[see][and references therein]{komo08}, which may also lead to generally weaker optical emission from the reprocessing process. Thus, the weakness or absence of this excess variation component may also contribute, at least partly, to the overall lower variability amplitudes in NLS1- compared to BLS1-type AGNs.

If the dominant mechanism for the long-term optical/UV variability is related to accretion disk instability, we would expect that the characteristic variability timescales inferred from structure function are different between NLS1- and BLS1-type AGNs. The typical physical timescales associated with the accretion disk scale with the black hole mass. However, in our analysis the variability timescales inferred from the exponential fits of the structure function do not show statistically significant difference between the two types of AGNs given their large uncertainties. This may be ascribed, at least partly, to the relatively small range of  black hole masses in our sample. We note that in the analysis of SDSS Stripe-82 data for massive quasars, \citet[][]{macl10} find  the characteristic timescale correlates with the black hole mass with a power law index of 0.21$\pm$0.07. While AGNs with relatively smaller black hole mass, such as the NLS1-type AGNs are not included in their sample.
The property of the characteristic timescales of AGN variability may be examined by using large samples with a wider black hole mass range and  monitored over longer time spans than the currently available data, such as the upcoming Large Synoptic Survey Telescope.

\subsection{Implication for the nature of NLS1-type AGNs}

We have demonstrated that, on all timescales observed so far,  NLS1-type AGNs show systematically less optical/UV variations compared to BLS1-type AGNs. This is in agreement with the dependence of the variability on the Eddington ratio for BLAGNs found in our previous paper \citep[][]{ai10}. In fact, as we argued, the optical/UV variability may be considered as a new EV1 parameter, with NLS1s lying at one extreme end. This property gives interesting insight into the nature of NLS1-type AGNs.
Currently there are three scenarios regarding the nature of NLS1-type AGNs.
The widely accepted model invokes less massive black holes
accreting at high rates compared to BLS1s \citep[e.g.][]{mine00, sule00,wang03}.
Alternatively, the narrow Balmer line width can also be explained
by a disk-like low-ionization BLR close to face-on
(the ``orientation scenario'', e.g.,\ Osterbrock \& Pogge 1985, Collin \& Kawaguchi 2004),
or by a more distant BLR compared to normal BLS1s
with a similar black hole mass or nuclear luminosity (``distant
BLR scenario'', Wandel \& Boller 1998).
Although face-on inclination is found in a small number of
extreme radio-loud NLS1s showing blazar-like properties
(Yuan et al.\ 2008, abdo et al. 2009, Gu \& Chen 2010),
comparative studies of radio properties between NLS1 and BLS1
disfavor the orientation scenario in general (Zhou \& Wang 2002;
Komossa et al. 2006; Zhou et al. 2006).
Other lines of evidence, such as polarization properties and correlations involving
narrow-line luminosity, suggest that orientation at most plays a
secondary role in explaining NLS1 phenomenon (Komossa 2008 and
references therein).

Neither the ``orientation'' nor ``distant BLR'' scenario
can explain  naturally
the extension of EV1 correlations involving variability.
For the ``distant BLR'' scenario, it is hard to link the size of the BLR
to the optical/UV continuum variability, that is local to the accretion disk.
Also, since the optical continuum emission is believed to come from an accretion disk,
as long as the disk is directly seen, variability amplitude should not be
dependent on the orientation of the disk.
Optical/UV variability may be caused by  obscuration by the outer part of
the disk; however, a lack of periodic variability corresponding to
the disk rotation has ruled out this possibility.
(e.g., Peterson \& Bentz 2006; Wold et al. 2007).
In comparison, the commonly accepted model for NLS1s
with smaller \mbh\ and higher \redd\ is favored.
Their weak variability fits more naturally into the
scenario that variability amplitude is governed by \redd, as suggested in our previous paper.

\section{Summary}
We present the ensemble variability property of a sample of NLS1-type AGNs
that have multi-epoch photometric observations in the SDSS Stripe-82 region.
For NLS1-type AGNs the variability was hitherto poorly explored as a class
due to both the limited sample sizes and few data available.
As a direct comparison a control sample of BLS1-type AGNs is also compiled.
We introduce a novel parametric parametric maximum-likelihood method
in variability studies, to quantify intrinsic variability amplitudes and, for the first time, their confidence
intervals, which improves upon the commonly adopted methods in the literature.
This method makes full use of the measurement error of each data point,
and has the advantage of being sensitive to low-level variations.

We find that the majority of the NLS1-type AGNs are in fact variable
on timescales from about several days to a few years,
e.g., more than 80\% of the objects varied at levels $\geq$ 0.05\,mag (standard deviation)
in the SDSS $u$ and $g$ bands. NLS1-type AGNs have systematically smaller variability
compared to BLS1-type AGNs. This is consistent with the previously found
anti-correlation of AGN variability with the Eddington ratio
\citep[][]{wilh08,ai10,macl10}.
We present the light curves and some of the key parameters for
22 NLS1-type AGNs that show relatively
large variations (standard deviation $\sigma> 0.1$\,mag in the r-band),
which may deserve further  monitoring observations.
It is also found that, similar to BLS1-type AGNs (including quasars),
the optical/UV variability is wavelength--dependent---the shorter wavelength,
the larger variations.
We derive,  for the first time,
the ensemble structure function of variability for
NLS1-type AGNs that span nearly three orders of magnitudes
in time lags (from a few days to  $\sim10^3$ days).
The structure functions of NLS1- and BLS1-type AGNs show remarkably similar profiles on time lags $\ga$ 10 days, with the former having smaller amplitude at a given time lag.
The structure function can be approximated by either a power law or an exponential function, though the latter seems to give a better description for NLS1-type AGNs. On time scales of $\la$ 10 days, the variability amplitudes of BLS1-type AGNs remain somewhat higher than those extrapolated from the power law structure function, indicating the presence of some excess variations. However, this is not the case for NLS1s. We suggest this excess variation component to be likely arising from reprocessing of the X-rays, which might be more important in BLS1- than in NLS1-type AGNs.

In conclusion, NLS1- and BLS1-type AGNs share very similar properties of long-term
optical/UV variability in many ways, such as
the wavelength and time lag dependence of variability.
This suggests that most likely the same physical mechanisms
are at work to produce the long-term optical/UV variability in both types,
in spite of the possibly different modes of their accretion flows.
Future photometric monitoring programs for larger samples of AGN
spanning wide parameter ranges, with a large dynamic range of
sampling rate, are needed to address the questions related to the optical/UV
variability, such as the behavior at time lags less than a week and the characteristic variability timescales.

\acknowledgments

This work is supported by NSFC grants 11033007, 11103071, the National Basic
Research Program of (973 Program) 2009CB824800, 2007CB815405.
Funding for the SDSS and SDSS-II was provided by the Alfred P. Sloan Foundation, the
Participating Institutions, the National Science Foundation, the U.S. Department of Energy,
the National Aeronautics and Space Administration, the Japanese Monbukagakusho,
the Max Planck Society,
and the Higher Education Funding Council for England.
The SDSS Web Site is http://www.sdss.org/. Yanli Ai acknowledges the support from the LAMOST Postdoctoral Fellowship Program.



\clearpage

\begin{figure}
\nonumber
\centering\includegraphics[scale=1]{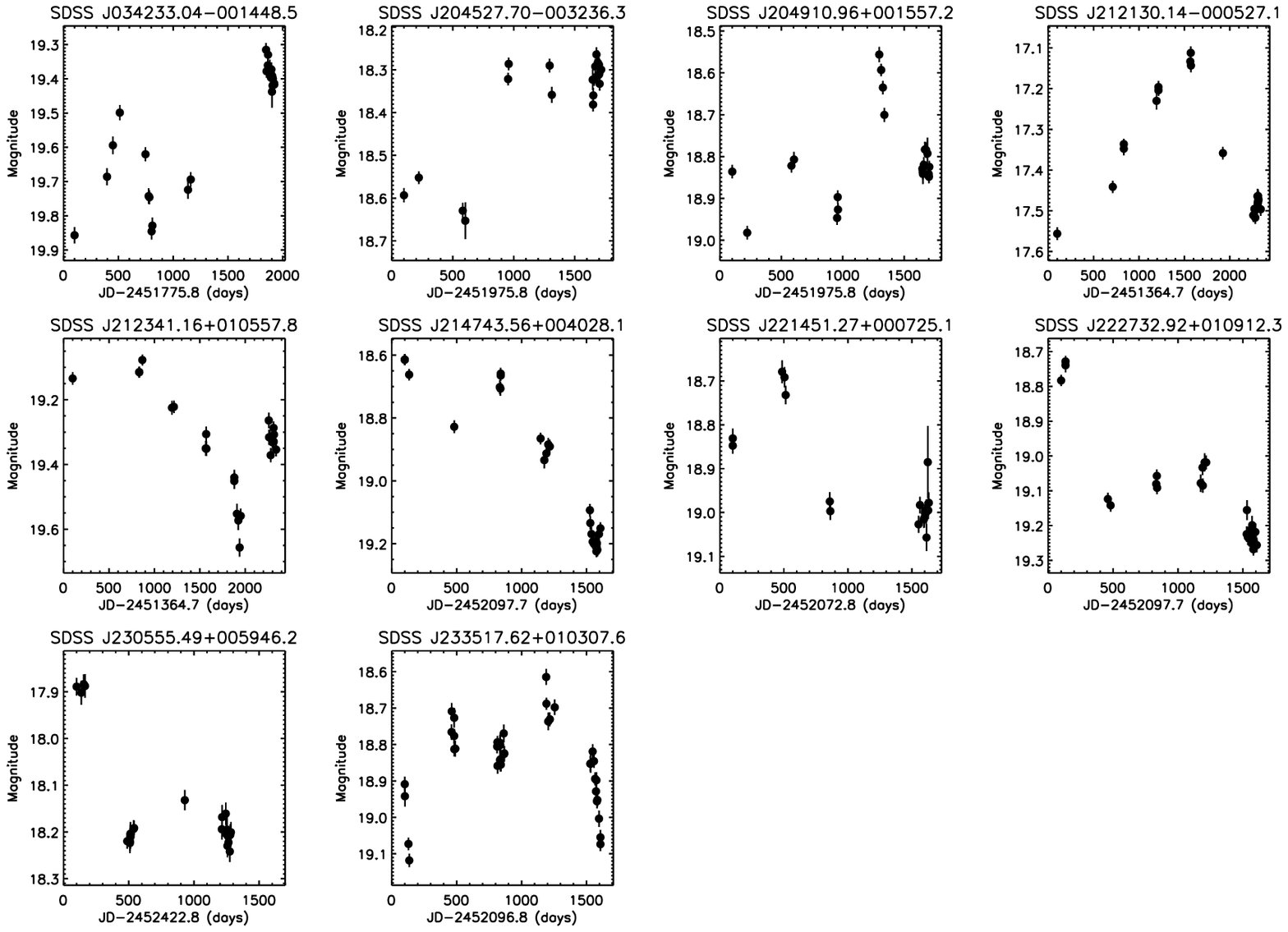}\par{Figure\,\ref{lightc_22} ---  \it{continued}}
\end{figure}

\clearpage

\begin{table}
\scriptsize \centering \caption{Fraction of variable objects and averaged magnitude of variations. \label{var_fraction}}
\begin{tabular}{ccccccccc}
\hline \hline & \multicolumn{4}{c}{NLS1-type AGNs} & \multicolumn{4}{c}{BLS1-type AGNs}  \\
\hline
bands & $P(\chi^{2}\mid\nu)\leq0.001^a$ & $\sigma_{m}\geq0.05^b$ & $\sigma_{m}\geq0.10^c$ &
$\langle\sigma_{m}\rangle^d$ &
$P(\chi^{2}\mid\nu)\leq0.001^a$ & $\sigma_{m}\geq0.05^b$ & $\sigma_{m}\geq0.10^c$ &
$\langle\sigma_{m}\rangle^d$ \\
\hline
u & 85.4 & 89.1 &53.4 & 0.11 & 93.5 & 90.6  &71.5 & 0.14 \\
g & 98.3 & 85.4 &44.8 & 0.10 & 99.1 & 95.3  &60.1 & 0.13 \\
r & 98.3 & 76.2 &41.3 & 0.09 & 99.1 & 90.6  &56.9 & 0.12 \\
i & 100. & 74.5 &39.6 & 0.08 & 97.2 & 89.7  &49.6 & 0.10 \\
z & 81.3 & 69.4 &22.4 & 0.07 & 85.0 & 80.3  &38.2 & 0.09 \\
\hline
\end{tabular}
\begin{flushleft}
$^a$ Fraction of sources with $P(\chi^{2}\mid\nu)\leq0.001$;
$^b$Fraction of sources with $\sigma_{m}\geq0.05$;
$^c$Fraction of sources with $\sigma_{m}\geq0.10$;
$^d$Mean $\sigma_{m}$ in each band
\end{flushleft}
\end{table}


\begin{table}
\scriptsize \caption{NLS1-type AGNs with $\sigma_r > 0.1$ mag
\label{NLS1_examp}}
\begin{tabular}{cccccccccc}
\hline \hline SDSS Name & Redshift &  FWHM(H$\beta$) &  log$\lambda L_{\lambda 5100}$  &  R$_{4570}$  &
$\langle r \rangle$ &
$\sigma_{g}$ & $\sigma_{r}$ &
max$\Delta$ g & max$\Delta$ r \\
(1) & (2) & (3) & (4) & (5) & (6) & (7) & (8) & (9)& (10) \\
\hline
J003431.74-001312.7&0.381&1701$\pm$31&44.47&0.27$\pm$0.01&18.14&0.11$\pm$0.05&0.10$\pm$0.04&0.37&0.33 \\
J010737.01-001911.6&0.737&1853$\pm$57&45.10&0.78$\pm$0.04&18.34&0.19$\pm$0.08&0.18$\pm$0.08&0.56&0.59 \\
J011712.82-005817.4&0.485&1987$\pm$65&44.51&0.49$\pm$0.03&19.24&0.15$\pm$0.05&0.14$\pm$0.04&0.49&0.47 \\
J012824.21+001925.2&0.419&1552$\pm$132&44.13&0.15$\pm$0.06&19.70&0.16$\pm$0.05&0.17$\pm$0.05&0.55&0.53 \\
J013509.51+000252.0&0.744&1890$\pm$129&44.73&0.96$\pm$0.12&18.99&0.17$\pm$0.07&0.15$\pm$0.06&0.49&0.43 \\
J025501.18+001745.4&0.359&1636$\pm$120&44.04&1.07$\pm$0.14&19.41&0.17$\pm$0.05&0.11$\pm$0.03&0.59&0.40 \\
J025910.39-002239.8&0.360&2083$\pm$52&44.52&0.66$\pm$0.02&18.20&0.16$\pm$0.06&0.15$\pm$0.06&0.51&0.40 \\
J030335.76+004145.0&0.669&1999$\pm$83&44.87&0.13$\pm$0.02&19.24&0.15$\pm$0.04&0.14$\pm$0.04&0.59&0.57 \\
J031022.11+004130.0&0.655&1972$\pm$203&44.60&1.44$\pm$0.23&19.74&0.12$\pm$0.03&0.10$\pm$0.03&0.46&0.40 \\
J031427.46-011152.3&0.387&1709$\pm$22&44.59&0.50$\pm$0.01&18.15&0.14$\pm$0.05&0.14$\pm$0.06&0.54&0.50 \\
J031532.28+005503.6&0.487&2059$\pm$165&44.12&0.00$\pm$0.05&19.74&0.23$\pm$0.08&0.24$\pm$0.09&0.98&0.87 \\
J031630.78-010303.6&0.368&1174$\pm$87&44.28&0.79$\pm$0.07&18.73&0.15$\pm$0.04&0.12$\pm$0.03&0.55&0.41 \\
J034233.05-001448.5&0.695&1502$\pm$118&44.54&0.53$\pm$0.18&19.52&0.16$\pm$0.06&0.17$\pm$0.06&0.49&0.54 \\
J204527.70-003236.2&0.296&1473$\pm$53&44.20&1.13$\pm$0.04&18.37&0.14$\pm$0.06&0.11$\pm$0.05&0.44&0.38 \\
J204910.97+001557.3&0.362&1556$\pm$41&44.17&0.25$\pm$0.03&18.80&0.12$\pm$0.06&0.10$\pm$0.04&0.47&0.42 \\
J212130.14-000527.1&0.585&1898$\pm$39&45.32&0.95$\pm$0.03&17.39&0.15$\pm$0.07&0.13$\pm$0.05&0.40&0.44 \\
J212341.16+010557.9&0.389&1473$\pm$52&44.23&0.75$\pm$0.07&19.33&0.16$\pm$0.06&0.14$\pm$0.05&0.66&0.57 \\
J214743.56+004028.2&0.643&1987$\pm$72&44.94&0.80$\pm$0.03&18.98&0.22$\pm$0.08&0.22$\pm$0.08&0.68&0.61 \\
J221451.27+000725.1&0.573&1780$\pm$87&44.69&1.01$\pm$0.08&18.93&0.09$\pm$0.04&0.11$\pm$0.05&0.32&0.37 \\
J222732.93+010912.3&0.546&1472$\pm$46&44.61&0.68$\pm$0.04&19.11&0.16$\pm$0.06&0.15$\pm$0.05&0.58&0.53 \\
J230555.49+005946.2&0.719&2161$\pm$47&45.17&0.44$\pm$0.02&18.14&0.11$\pm$0.04&0.11$\pm$0.05&0.33&0.35 \\
J233517.62+010307.7&0.624&2095$\pm$109&44.66&0.01$\pm$0.03&18.84&0.10$\pm$0.03&0.10$\pm$0.03&0.51&0.50 \\
\hline
\end{tabular}
\tablecomments{
Col.(1): SDSS name;
Col.(2): redshift;
Col.(3): H$\beta$ linewidth (km\,s$^{-1}$);
Col.(4): monochromatic luminosity at 5100\,{\rm \AA} (ergs\,s$^{-1}$);
Col.(5): intensity ratio of the \feii\ multiplets to H$\beta$;
Col.(6): mean magnitude in $r$ band;
Col.(7)-(8): best estimated variability amplitudes and corresponding errors at 90\% confidence level from maximum-likelihood method;
Col.(9)-(10): maximum magnitude changes
}
\end{table}

\begin{table}
\scriptsize \centering \caption{Results of model fits to the structure functions at $\Delta\tau >$ 10 days \label{SF_model_fits}}
\begin{tabular}{ccccc}
\hline \hline Band& \multicolumn{2}{c}{NLS1-type AGNs} & \multicolumn{2}{c}{BLS1-type AGNs} \\
\mytoprule
\multicolumn{5}{c}{Power Law} \\
\mybottomrule
 & $\Delta\tau_{0}$(d)& $\beta$& $\Delta\tau_{0}$(d)& $\beta$ \\
\hline
u&7.3$\times 10^{4}$ &0.33$\pm$0.01 &1.7$\times 10^{4}$ &0.36$\pm$0.01 \\
g&6.3$\times 10^{4}$ &0.36$\pm$0.01 &2.2$\times 10^{4}$ &0.37$\pm$0.01 \\
r&5.4$\times 10^{4}$ &0.40$\pm$0.01 &3.1$\times 10^{4}$ &0.35$\pm$0.01 \\
i&8.1$\times 10^{4}$ &0.39$\pm$0.01 &4.3$\times 10^{4}$ &0.36$\pm$0.01 \\
z&3.6$\times 10^{5}$ &0.32$\pm$0.02 &5.7$\times 10^{4}$ &0.37$\pm$0.01 \\
\hline
$\Delta\lambda_{1900-2500\,\AA}$&1.4$\times 10^{6}$&0.30$\pm$0.02& 2.1$\times10^{4}$&0.34$\pm$0.01\\
$\Delta\lambda_{2500-3300}\,\AA$&6.6$\times 10^{4}$&0.36$\pm$0.01& 2.9$\times10^{4}$&0.35$\pm$0.01\\
$\Delta\lambda_{3300-4200}\,\AA$&5.7$\times 10^{4}$&0.38$\pm$0.01& 2.6$\times10^{4}$&0.36$\pm$0.01\\
$\Delta\lambda_{4200-5500}\,\AA$&6.5$\times 10^{4}$&0.40$\pm$0.01& 3.2$\times10^{4}$&0.38$\pm$0.01\\
$\Delta\lambda_{5500-7100}\,\AA$&3.7$\times 10^{5}$&0.33$\pm$0.02& 8.6$\times10^{4}$&0.34$\pm$0.01\\
\mytoprule
\multicolumn{5}{c}{Exponential} \\
\mybottomrule
 & $\tau$(d)& $SF_{\infty}$& $\tau$(d)& SF$_{\infty}$ \\
\hline
$\Delta\lambda_{1900-2500}\,\AA$&217$\pm$28&0.21$\pm$0.01&282$\pm$24&0.32$\pm$0.01\\
$\Delta\lambda_{2500-3300}\,\AA$&336$\pm$33&0.21$\pm$0.01&309$\pm$21&0.28$\pm$0.01\\
$\Delta\lambda_{3300-4200}\,\AA$&404$\pm$40&0.21$\pm$0.01&343$\pm$21&0.28$\pm$0.01\\
$\Delta\lambda_{4200-5500}\,\AA$&562$\pm$58&0.20$\pm$0.01&424$\pm$27&0.26$\pm$0.01\\
$\Delta\lambda_{5500-7100}\,\AA$&410$\pm$67&0.14$\pm$0.01&401$\pm$39&0.21$\pm$0.01\\
\hline
\end{tabular}
\end{table}

\clearpage

\appendix

\section{Re-calibration of Supernova Survey data}
\label{sect:calibration}

Re-calibration of SN data is performed in five
bands separately in a field-by-field (100\,arcmin$^{2}$ patches) manner, and only fields with ``good"
image quality (flagged as ``good'' or ``acceptable'')
were used for photometric calibration or light curve construction in our work.
Photometric conditions can be regarded to be
unchanged across the field. For each of the SDSS SN survey fields,
we only use the most overlaying DR5 field to do calibration. In the
case of more than one overlapped DR5 fields, we use the one with the
larger or largest overlapping area. In each field and each band,
`standard stars' with good measurements (high quality photometry)
are selected with the following criteria:

1. Sources must be `star'.

2. The processing flags BRIGHT, SATURATED
   BLENDED, EDGE, NOPROFILE, INTERP\_PROBLEMS, and DEBLEND\_PROBLEMS
   are not set in any band\footnote{See
   http://www.sdss.org/dr7/products/catalogs/flags.html.}.

3. Sources must be brighter than the magnitude limits
   which are 22.0, 22.2, 22.2, 21.3, 20.5 in $ugriz$ bands.

4. Potentially variable sources were excluded, including QSOs,
   known variable stars, etc. Here QSO identification was based on SDSS quasar
   catalog. III. \citep[][]{schn05}, and variable stars based on the
   General Catalog of Variable Stars, 4th Ed. (GCVS4) \citep[][]{khol98}.

The selected stars were then cross-identified between the two most
overlaying fields with a matching radius of $1^{''}$. Any stars with
no match or more than one matches within this radius were discarded.
We computed various photometric statistics, such as mean magnitudes
and mean standard errors (weighted by statistical errors), and rms
scatter of the magnitude difference of the matched stars. Stars with
magnitude difference more than two times of the rms scatter were
rejected in order to avoid variable stars, mismatched stars, or bad
data points, etc. The above procedure was repeated until no more
star was excluded. The final number of calibration stars is
typically $\sim 50-200$ for each pair of fields, which is large
enough to do calibration with an acceptable uncertainty.
For each of the fields,
the differential magnitudes of the calibrating stars
between the SN survey and DR5
observations ($\Delta m=m_{\rm SN}- m_{\rm DR5}$) are
calculated, and their weighted mean is set to zero,
$\langle \Delta m \rangle=0$.
In this way the zero-point offsets of the SN survey photometry are determined,

The internal uncertainties of the calibrated zero-points of the SN survey
photometry with respect to the reference DR5 data,
as measured by the standard deviation of the
weighted mean of $\Delta m$, i.e., $\sigma_{\langle \Delta m \rangle}$,
are less than 0.01 mag for all the fields.
However, the SDSS SN survey data are calibrated using different fields
of SDSS-I DR5 data,  which have  typical systematic photometric zero-point
errors ($\sigma_{\rm zpt}^{DR5}$) of 0.01\,mag for the $g, r$, and
$i$ bands, 0.02\,mag for the $z$ band,
and 0.03\,mag in the $u$ band \citep[][]{Ivez04}.
Thus the total systematic uncertainties of the SN survey photometry
are estimated to be the combination in quadrature of
the above two terms, i.e.\
$\sqrt{(\sigma_{\langle \Delta m \rangle})^2 +(\sigma_{\rm zpt}^{DR5})^2}$.
The overall photometric errors are then
a contribution of both the systematic and statistical errors.
The latter are given in the SN survey catalogs and
range from a few percent to 0.1 or higher toward faint magnitudes,
depending on the bands.
The final photometric errors of the calibrated SN survey data have a median of
$\approx0.03$\,mag in the $g$, $r$, and $i$ bands, and somewhat higher
($\approx 0.04$\,mag) in the $u$ and $z$ bands.
They are comparable to those of the DR5 data observed in SDSS-I.

\begin{figure}
\includegraphics[scale=0.9]{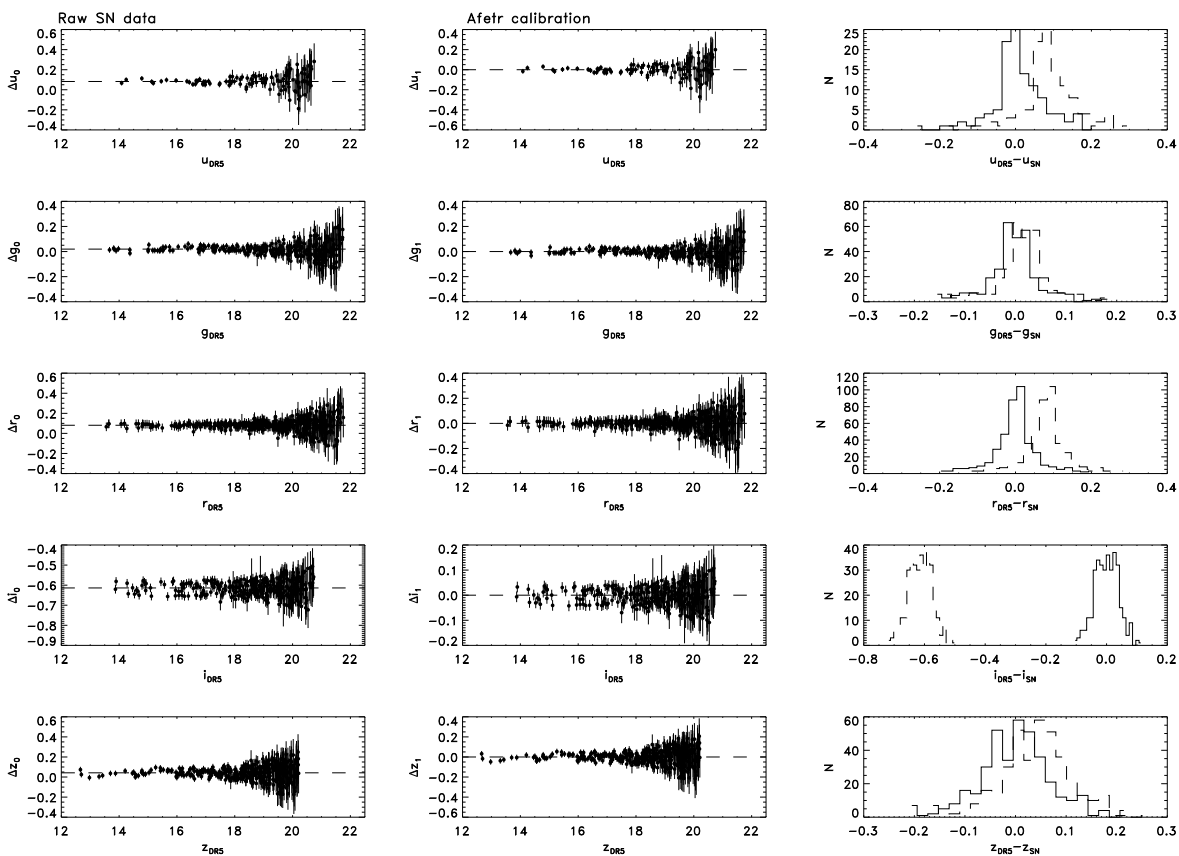}
\centering \caption{Illustration of the zero-point calibration for
one field in SDSS Supernova survey, with the cross-identified star
from the most overlaying filed in DR5. The dots in the left panels
represent magnitude differences between the DR5 and the original SN
magnitudes, as a function of the DR5 magnitude. The dashed line
shows the mean magnitude differences weighted by statistical errors.
The middle panels show the magnitude differences after our
calibration. The histograms of the magnitude differences before
(dashed line) and after (solid line) calibration are compared in the
right panels. \label{calibration}}
\end{figure}

As an example, the SN data of one field before and after
calibration are illustrated in Fig. \ref{calibration}. Actually our
photometric accuracy is much better than the rms scatter shown in
Fig. \ref{calibration} because most of the objects in our samples
are brighter than 20.0 mag in $r$ band.

\clearpage

\clearpage
\section{Amplitude of Variability and the Maximum-likelihood method }
\label{sect:method}

A common practice as used in the literature is to estimate the
contribution of the  errors of measurement to the observed scatter,
and then subtract it from the latter \citep[][]{vand04, sesa07}. The observed rms scatter $\Sigma$ is calculated as
\begin{equation}\label{eq2}
\Sigma = \sqrt{\frac{1}{n-1} \sum\limits_{i=1}^{N}(m_{i} - \langle m
\rangle)^2},
\end{equation}
where  $m_{i}$ are observed magnitudes of N observations
and $\langle m \rangle$ is their weighted mean.
The intrinsic variation amplitude \sgmint\ can be estimated as
\begin{equation}\label{eq4}
\sigma_{\rm method-1}=\left\{%
\begin{array}{ll}
    (\Sigma^2-\xi^2)^{1/2}, & {\rm if~\Sigma > \xi,}\\
    0, & {\rm otherwise.} \\
\end{array}%
\right.
\end{equation}
where $\xi$ is the term representing the amount of scatter
caused by measuremental errors.
The estimation of $\xi$
has some variances among the literature.
For example, \citet[][]{sesa07} used the mean photometric errors
of SDSS observations as a function of magnitude which was fitted
from the data assuming most stars are not variable;
a theoretically expected  $\xi(m)$ as a function of magnitude
was also given in \citet[][]{stra01}.
Here we estimate $\xi$  as the mean square value of the errors
$\xi_{i}$ associated with the individual magnitude  $m_{i}$, i.e.
\begin{equation}\label{eq3}
\xi^2 = \frac{1}{N} \sum\limits_{i=1}^{N} \xi_{i}^{2}.
\end{equation}
This estimation is similar that used in Rodiguez-Pascual et al. (1997),
which was in the flux domain rather than magnitude though.

In this paper, we introduce
a new method---the maximum-likelihood method---to
quantify \sgmint\ and its uncertainty,
in the line of the parametric statistics approach.
There are two assumptions involved in this method.
Firstly, we assume that, for a given measurement,
the probability function of the
random fluctuations in the measurement follows a Gaussian
distribution\footnote{This is supported
by the fact that for stars (mostly non-variable)
repeatedly observed in the SDSS with similar photometric errors,
their magnitude distributions can be well described as a Gaussian
({Ivezi\'c et al. (2003)}).
},
with a standard deviation of $\sigma_{r}$,
where $\sigma_{r}$ is the photometric error for this measurement.
Furthermore, for simplicity,
we approximate the intrinsic distribution of the magnitude of a variable object
as a Gaussian\footnote{
We consider this as a reasonable approximation, as
it is found from our data that, for some objects with a much larger
amplitude of intrinsic variations than their typical photometric errors
(i.e.\ the latter is negligible),
their magnitude distributions can be reasonably well described by a Gaussian.
},
with a mean $\langle m \rangle$ and a standard deviation
\sgmint\---both parameters are to be estimated.
\sgmint\  gives a measure of the amplitude of variability,
and is expected to be consistent with zero in case of no variation.

The final probability function of observing an object to have
a magnitude $m$ with a photometric uncertainty is then
\begin{equation}\label{eq5}
p(m) = \frac{1}{\sqrt{2\pi}\varphi} \exp\left[-\frac{(m-\langle m
\rangle)^2}{2\varphi^2}\right]
\end{equation}
Here $\varphi = (\sigma_{method-2}^{2}+\sigma_{r}^{2})^{1/2}$.
Suppose, for a given object, a set of values for $m$ are observed,
$m_{1}$, ..., $m_{N}$, with the corresponding (different) accuracies
of the measurements which are characterized by the standard
deviations $\sigma_{r1}$, ..., $\sigma_{rN}$. The likelihood
function, $\L(\langle m \rangle, \sigma_{method-2})$, of this data
set is
\begin{eqnarray}\nonumber\textstyle
\prod\limits_{i=1}^{N}
\frac{1}{\sqrt{2\pi}(\sigma_{method-2}^{2}+\sigma_{ri}^{2})^{1/2}}
\exp\left[-\frac{(m_{i} -\langle m
\rangle)^2}{2(\sigma_{method-2}^2+\sigma_{ri}^{2})}\right]
\end{eqnarray}
The maximum-likelihood estimates of $\langle m \rangle$ and
$\sigma_{method-2}$ for a given object are obtained by minimizing $S
= -2\ln\L$

This method can also be used to find the confidence intervals for
the interesting parameters, by applying the standard
$\Delta\chi^{2}$ techniques to the above S-function (see e.g.\  Avni
1976) . Suppose the parameter vector $\theta$ has two components,
and the first ``interesting component'' $\theta_{i}$ consists of $q$
parameters ($\theta_{1},...,\theta_{q}$), which need to be estimated
simultaneously without considering the second ``uninteresting
component'' $\theta_{u}$ \citep[see][]{avni76}. The confidence interval at
a probability level $\alpha$ for $\theta_{i}$ can be computed by
finding the set of all values of $\theta_{i}$ such that
\begin{equation}
S(\theta_{i},minimize~over~\theta_{u}) - S_{min} \leq \Delta(q,
\alpha),
\end{equation}
where $\Delta(q, \alpha)$ follow a $\chi^{2}$ distribution with a
degree of freedom $q$. In our case, there are two interesting
parameters ($q=2$), and the corresponding $\Delta(q,\alpha)$=2.30,
4.61, 9.21 for probability levels  $\alpha=68\%, 90\%, 99\%$,
respectively.

\begin{figure}
\includegraphics[scale=0.65]{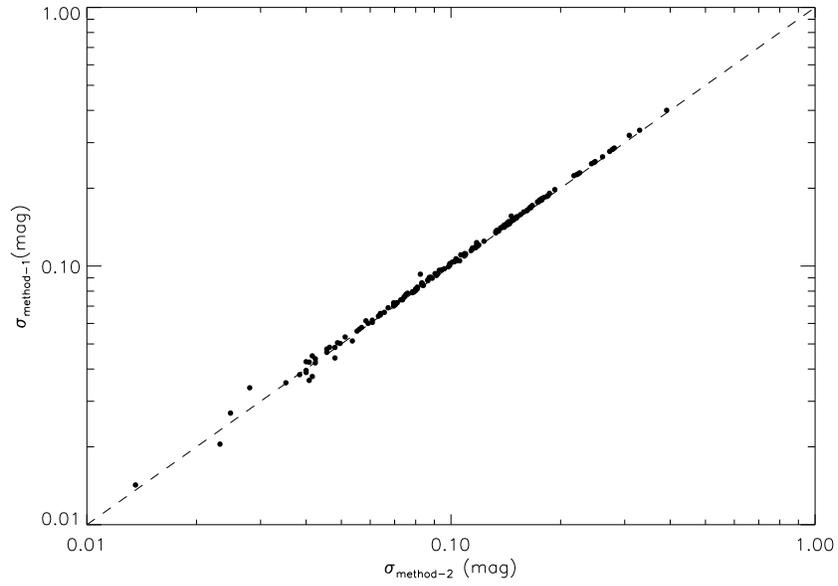}
\centering \caption{Comparison of the estimated variability
amplitudes with the two methods in $r$ band.
The other four bands give similar results. \label{compare_method}}
\end{figure}

Intrinsic variability amplitudes estimated via the above two methods
are compared in Fig.\,\ref{compare_method} and are found to be in
excellent agreement.

\end{document}